\documentstyle[preprint,prb,aps]{revtex}
\begin{document}
\author
{
J.P. Badiali
 \\
\vspace{0.5cm}
\textit{LECA, ENSCP-Universit\'e Pierre et Marie Curie,}\\
\textit{4 Place Jussieu, 75230 Paris Cedex 05, France}\\
}
\title { \bf{ENTROPY: FROM BLACK HOLES TO ORDINARY SYSTEMS.                 
                                    }}\maketitle

\begin{abstract}
Several results of black holes thermodynamics can be considered as firmly 
founded and formulated in a very general manner. From 
this starting point we analyse in which way these results may give us the 
opportunity to gain a better understanding in the thermodynamics of ordinary 
systems for which a pre-relativistic description is sufficient. First, we 
investigated the possibility to introduce an alternative definition of the 
entropy basically related to a local definition of the order in a spacetime 
model rather than a counting of microstates. We show that such an 
alternative approach exists and leads to the traditional results provided an 
equilibrium condition is assumed. This condition introduces a relation 
between a time interval and the reverse of the temperature. We show that 
such a relation extensively used in the black hole theory, mainly as a 
mathematical trick, has a very general and physical meaning here; in 
particular its derivation is not related to the existence of a canonical 
density matrix. Our dynamical approach of thermodynamic equilibrium allows 
us to establish a relation between action and entropy and we show that an 
identical relation exists in the case of black holes. The derivation of such 
a relation seems impossible in the Gibbs ensemble approach of statistical 
thermodynamics. From these results we suggest that the definition of entropy 
in terms of order in spacetime should be more general that the Boltzmann one 
based on a counting of microstates. Finally we point out that these results 
are obtained by reversing the traditional route going from the 
Schr\"{o}dinger equation to statistical thermodynamics.

PACS number:03.65Ca, 05.30-d, 05.70-a, 47.53+n .
\end{abstract}
\vspace{0.5cm}

\section{Introduction}
After the discovery of the Hawking radiation (\cite{hawking1}) it became 
clear that the similarity observed by Bekenstein (\cite{bekenstein1}) 
between the four laws of black hole mechanics and the ones of standard 
thermodynamics is much more than a simple analogy but represents a result 
having a deep significance. For a given black hole, the same 
values for temperature, $T_{BH}$, and entropy, $S_{BH}$ have been obtained 
from different approaches. Consequently, we may consider the black 
hole thermodynamics as firmly established; one exception, may be, concerns 
the derivation of the second law of thermodynamics that is replaced by the 
so-called general second law (\cite{wald1}). The main question emerging from 
these results is relative to the physical origin of the black hole entropy. 
It is implicitely assumed that the physical content of $S_{BH}$ can be 
understood, at least to some extent, from the usual ingredients of classical 
physics. This leads to search a relation between $S_{BH}$ and a counting of 
microstates. After thirty years of intensive works in that direction we have 
still to deal with questions relative to the nature and the location in 
space of these microstates (\cite{bekenstein2},\cite{sork0}).\\ 
In this paper we reverse this problematic: Instead of trying to understand 
something concerning the black hole physics starting from the results 
obtained in ordinary physics, we try to see if it is possible to learn 
something about ordinary systems starting from results obtained in the 
domain of black holes. By ordinary systems we mean those for which a 
pre-relativistic description is sufficient. Obviously, we cannot expect a 
strict mapping between the thermodynamic properties of ordinary systems and 
those of systems including black holes for which there are fundamental 
singularities, presence of horizons and existence of a holographic 
principle. Nevertheless, as we shall see, it is possible to reformulate some 
results obtained in the black hole domain in a so general manner that it 
appears natural to ask wether it exists similar results in the case of 
ordinary systems.\\ 
It is well known that the Hawking tenperature, $T_{H}$, of a Schwarzschild 
black hole of mass $M$ measured at a large distance from the hole is given 
by $T_{H} = \frac{\hbar}{8 \pi M}$. In the 
original derivation of this result Hawking (\cite{hawking1}) used techniques 
of quantum field theory on a given classical curved background spacetime. 
For this purpose it was useful to work in the " real Euclidean section" of 
the Schwarzschild geometry, in which the time is rotated to its imaginary 
value and the thermal Green function is periodic with a period $\beta \hbar 
= \frac{\hbar}{k_{B} T_{H}}$ (see for instance (\cite{fulling}) for a 
mathematical analysis of this result). Later, Gibbons and Hawking 
(\cite{gibbons}) were able to deduce thermodynamics of black hole from 
statistical mechanics. The main steps in their derivation are the following. 
A canonical partition function, $Z_{GH}$, is introduced using a path 
integral in which the action is the one of the gravitational field and 
the time on which the action is defined is $\beta \hbar$. Then a zero loop 
approximation is performed leading to a free energy defined by $F_{GH} = - 
k_{B}T_{H} \ln{Z_{GH}}$ directly proportional to the action and 
from the usual thermodynamic relations it is possible to calculate the 
entropy. For four different metrics, it has been 
shown that the results derived in this way agree with previous derivations 
although the methods are totally different. Thus the entropy derived from 
the spacetime properties has a geometrical origin, it is not based 
explicitely on the counting of microstates although it coincides with the 
entropy derived from the thermodynamics. \\ It is tempting to retain from 
these results some general aspects that we may try to extend to ordinary 
systems. First, in parallel to the usual definition of entropy it might 
exist an alternative definition connected with the geometry of space-time 
and that we might establish without explicit reference with a counting of 
microstates. In what follows we shall see that 
such an alternative definition of entropy exists also for ordinary systems 
provided a spacetime model is introduced as suggested in (\cite{jpb1}). This 
spacetime is formed by a discrete manifold on which exists a structural  
relation that mimics the uncertainty relations and a given kinematics. An 
equilibrium condition introduces a time $\beta \hbar$, which is proportional 
to the reverse of the temperature. For this time our entropy is identical to 
the traditional one expressed in terms of path integral. Second, black hole 
theory suggests the existence of a relation between action and spacetime 
free energy leading to a relation between action and entropy. Hereafter we 
deduce a similar relation that is verified for black holes and for ordinary 
systems. For the simple case considered here, the result is quite general 
$i.e.$ it exists beyond the zero loop approximation. It is noteworthy that 
the existence of a relation between action and entropy has been suggested by 
Eddington a long time ago and reinvestigated later by de Broglie 
searching a relation between two quantities that are considered as 
relativistic invariants in restricted relativity (for a review in this 
domain see (\cite{debroglie})). \\ 
This paper is organized as follows. In 
Section 2 we summarize the properties of the spacetime model. In Section 3 
we introduce the path-entropy and other quantity giving a global information 
about the spacetime structure. In Section 4 we analyse the relation time / 
temperature from which we establish a link with ordinary thermodynamics. We 
point out the analogies but also the differences with the fouding of a 
similar relation existing in the case a thermal Green function defined in 
imaginary time that are commonly used in the black hole theory. In Section 4 
we consider the relation between action and entropy. In the last Section we 
give some concluding remarks.  

\section{Spacetime model} 
In his book with Hibbs, Feynman (\cite{feyn}) developped a given number of 
fundamental remarks concerning the derivation of the partition function in 
term of path-integral. He suggested a possible new foundation of statistical 
mechanics directly in terms of path integral as he did for quantum 
mechanics. In this context "directly" means without using the 
Schr\"{o}dinger equation. Note that this Feynman's conjecture implies to 
give a physical meaning to the paths and also to justify why one path 
integral is sufficient to describe statistical physics while only a product 
of path integral via the square of the amplitude of the wave function has a 
physical meaning in quantum physics. \\ One the most economical way to give 
a physical meaning to the paths consists in assuming the existence of a 
primarily discrete spacetime (\cite{jpb1}). Although this is probably not 
strictly needed we assume that the spacetime points $(t_{i},x_{i})$ are 
located on the sites of a regular lattice as in the chessboard model 
investigated in (\cite{feyn}). The spacetime structure is characterized by 
the existence of a relation between the elementary length $\Delta x$ and 
time interval $\Delta t$ corresponding to the lattice spacing. When a mass 
is introduced in this lattice we assume that $(\Delta x)^{2} / \Delta t 
=\hbar /m$, a relation mimicing the Heisenberg uncertainty relations. \\
We assume that the free motion in this spacetime model is as simple as 
possible. By definition, a path corresponds to a set of sites $(t_{i}, 
x_{i})$; the values of ${t_{i}}$ are such as $t_{i+1} > t_{i}$ whatever $i$ 
and the coordinate positions, $x_{i+1}$ is necessarily one of the nearest 
neighbours of $x_{i}$, thus a path corresponds to a random walk. \\
In this spacetime model we have a discrete manifold, the quantification 
appears via the relation between $\Delta x$ and $\Delta t$ and the 
kinematics is defined in terms of paths on which exist a causal relation.   
There is no metric as in the usual mathematical model of the general 
relativity. However, from the results obtained by Sorkin $et$ $al$ (see for 
instance (\cite{sork1}, \cite{sork2})) we know that a causal structure may 
determine almost all the information needed to specify the metric up to a 
multiplicative function called the conformal factor; the lack of conformal 
factor means that we have no quantitative measure for lengths and volume in 
spacetime. Our approach is reminiscent of the of causal sets theory on 
several points: we have a discrete spacetime, the paths defined above obey 
to causal relation and even the relation between $\Delta x$ and $\Delta t$ 
is not sufficient to fix a measure for lengths. As in the causal sets theory 
to define a length we have to say, for instance, how many spacetime points 
exists in a given volume. This can be done if, from a physical argument, we 
may fix the value of $\Delta x$, for instance. \\ 
In absence of such cutoff the limits $\Delta x, \Delta t \to {0}$ taken with 
 the constraint $(\Delta x)^{2} / \Delta t =\hbar /m$ leads to a continuous 
 diffusion process (\cite{itz}) for which 
the diffusion coefficient is $D = \hbar /2m$. In presence of an external 
potential, $u(t,x)$, this one is simply added to the diffusion equation 
(\cite{jpb1}). Using the Feynman-Kac formula, the fundamental solution, 
$q(t_{0},x_{0};t,x)$, of this new equation appears as a weighted sum of all 
the paths connecting the space-time points $(x_{0},t_{0})$ to $(x,t)$ ; we 
have \begin{equation} q(t_{0},x_{0};t,x) = \smallint {\cal D} x(t) \exp 
-\frac{1}{\hbar} A^{E}[x(t); t, t_{0}] \label{transfunc} \end{equation} 
where ${\cal D}x(t)$ means the measure for the functional integral and 
\begin{equation} A^{E}[x(t); t, t_{0}] =  \int\limits_{t_{0}}^{t}  
 [\frac{1}{2} m [\frac{dx(t')}{dt'}]^{2} + u(t',x(t'))] dt'. 
\label{action}  
\end{equation}
At this level it is important to underline several points. The paths 
are associated with processes that occur in real time. These processes are 
generated from the euclidean action $A^{E}[x(t);t,t_{0}]$ and 
they are such that, in average, there is no derivative $i.e.$ no velocity in 
the usual sense on the paths (\cite{jpb2}). The continuous limits $\Delta x, 
\Delta t \to {0}$ are useful to give a physical meaning to the processes 
but, at least in principle, the explicit calculation of the path integral 
can be performed with finite values of $\Delta x$ and $\Delta t$ that appear 
as natural cutoffs in the discretization procedure needed to calculate 
(\ref{transfunc}). The function $q(t_{0},x_{0};t,x)$ is not a density of 
probability although it verifies a Chapman-Kolmogorov law of composition. In 
what follows we will show that the function $q(t_{0},x_{0};t,x)$ is 
sufficient to describe a lot of spacetime properties. In particular we will 
introduce it in a functional over paths used to describe the order in 
spacetime.

\section{Path-entropy}
To define the order - or disorder - in spacetime we first adopt a local 
definition. Around a point $x_{0}$ we count the number of closed paths that 
we can form during a given time interval $\tau$. If there is only one possible 
path we can say that we have a perfect order, no fluctuation around this path 
is accepted. However, after introducing a given measure, some fluctuations 
can take place and we have to deal with a given number of acceptable paths. 
For this measure, we may associate the order in spacetime with this number 
of paths. The total order in our system will be obtained by summing the 
result of this procedure on all the points $x_{i}$ existing in the 
spacetime. This definition seems quite natural anytime we have to deal with 
processes occuring in a given spacetime. Of course such a definition is not 
unique but it is probably the simplest one. \\
By analogy with the thermodynamic entropy, which is defined for given values 
of internal energy and volume we consider that our spacetime system 
is prepared with a given energy $U$ and filled a volume $V$. We define a 
path-entropy by counting the number of paths for which the euclidean 
action that we note hereafter as $A^{E}[x(t); \tau]$, does 
not deviate too much from the action $\tau U$. In reference  
with the standard thermodynamics we define a path-entropy, $S_{path}$, 
according to  
\begin{equation} 
S_{path} = k_{B} \ln \smallint dx_{0} \smallint {\cal D}x(t) \exp{ - 
\frac{1}{\hbar} [A^{E}[x(t); \tau] - \tau U] }. 
\label{spath} 
\end{equation}
$S_{path}$ can be also rewritten as 
\begin{equation} 
S_{path} = \frac{k_{B} \tau}{\hbar} U + k_{B} \ln Z_{path} 
\label{entpath} 
\end{equation}  
with 
\begin{equation} 
Z_{path} = \smallint dx_{0} \smallint {\cal D}x(t) \exp{ - 
\frac{1}{\hbar} A[x(t); \tau]} = \smallint dx_{0}q(0,x_{0}; \tau, x_{0}) 
\label{zpath} 
\end{equation} 
in which $q(0,x_{0}; \tau, x_{0})$ corresponds to closed paths for which 
$t_{0} = 0$ and the time interval $\tau$. $Z_{path}$ is the total number of 
closed paths that we may count during $\tau$ irrespective the value of $U$. 
$S_{path}$ contains two external parameters $\tau$ and $U$ while $Z_{path}$ 
is only function of $\tau$. We may consider the dependence of $S_{path}$ 
versus these two parameters by considering the two derivatives : $\frac {d 
S_{path}}{dU}$ defined as $\frac {1}{T_{path}}$ and $\frac {d 
S_{path}}{d\tau}$. From the results given in (\cite{jpb1}) we have : 
\begin{equation} 
\frac{\hbar}{k_{B}} \frac{d S_{path}}{dU} = \frac{\hbar}{k_{B}} 
\frac{1}{T_{path}} = \tau + [ U -  (<u_{K}>_{path}+ <u_{P}>_{path})]\frac{d 
\tau}{dU} \label{Tpath} 
\end{equation}
and
\begin{equation} 
\frac{\hbar}{k_{B}} \frac{d S_{path}}{d\tau} = [ U -  
(<u_{K}>_{path}+ <u_{P}>_{path})] + \tau \frac {dU }{d \tau} 
\label{Spathto} 
\end{equation}
in which a relation between $U$ and $\tau$ is assumed. The average over 
paths that appear in (\ref{Tpath}) and (\ref{Spathto}) are defined according 
to 
\begin{equation}
<u_{P}>_{path} =  \frac{1}{Z_{path}} \smallint dx_{0} u(x_{0})q(0,x_{0}; 
\tau, x_{0})
\label{upath}
\end{equation}
and 
\begin{equation} 
\frac{m}{2} <(\frac{\delta x }{\delta t})^{2}>_{path} 
= \frac{\hbar}{2 \delta t} - <u_{K}>_{path} \label {kinpath} 
\end{equation}
in which $<u_{K}>_{path}$ is a well behaved function in the limit $\delta t 
\to 0$ moreover we have checked on examples that $<u_{K}>_{path}$ is just 
the usual thermal kinetic energy (\cite{jpb1}).\\
The quantities, $S_{path}$, $Z_{path}$, $T_{path}$ and $\frac{d 
S_{path}}{d\tau}$ are well defined, they gives us a global characteristic of 
the spacetime structure but none of them corresponds to a thermodynamic 
quantity. In the next Section we show that such a correspondence can be 
established for a given value of $\tau$ and we discuss the existence of a 
general relation between time and temperature. 

\section{Relation between time and temperature}
In (\ref{Tpath}) the sum $<u_{K}>_{path} + <u_{P}>_{path}$ is only dependent 
on the parameter $\tau$ and we may choose $\tau$ in such a way that the previous 
sum coincide with $U$, thus $\tau$ corresponds to a condition of thermal 
equilibrium. Now from (\ref{Tpath}) we conclude that the relation between 
$\tau$ and the temperature $T_{path}$ is $ \tau = 
\frac{\hbar}{k_{B}T_{path}}$ whatever the value of $\frac{d \tau}{dU}$. If 
we identify $T_{path}$ with the usual temperature we can see that $Z_{path}$ 
defined in (\ref{zpath}) becomes identical to the traditional partition 
function expressed in terms of path integral (\cite{feyn}) and thus we may 
recover all the results of thermodynamics. It is quite simple to verify that 
(\ref{Spathto}) is now reduced to $\frac{\hbar}{k_{B}} \frac{d 
S_{path}}{d\tau} = \tau \frac{dU }{d \tau}$ that we can rewrite as $dU = 
TdS$. Here we may interpret this relation as follows: if we increase the 
energy $U$ for the system preparation we increase the number of paths 
avalaible and therefore the entropy in spacetime. It has also been shown that 
$\tau$ is the time interval that we have to wait in order to relax the 
quantum fluctuations and to reach a thermal regime (\cite{jpb3}). Finally we 
may also derive the relation $\tau = \beta \hbar$ from the following 
heuristic argument. From standard thermodynamics we know that if a change of 
energy $\Delta U$ produces on a moving body a change of momentum $\Delta P$, 
the corresponding change of entropy $\Delta S$ is given by (\cite{giles}): 
\begin{equation} \Delta S = \frac{1}{T} \Delta U - (\frac{V}{T}) \Delta P 
\label{deltas} \end{equation} in which $V$ is the velocity of the mobile. 
From (\ref{deltas}) we may learn two different kinds of results. First, in 
terms of variations we may assume that $\Delta U$ results from quantum 
fluctuations and its estimation is $\Delta U = \frac{\hbar}{2 \tau}$. In (\ref{deltas})
we can write $V \Delta P$
 as $\Delta(\frac{1}{2} m V^{2})$ and if we 
assume that the quantum fluctuations lead to the thermal equilibrium we have 
$\Delta(\frac{1}{2} m V^{2}) = \frac{1}{2 \beta}$. Since the system is at 
equilibrium we must have no net change of entropy during these fluctuations 
$i.e.$ $\Delta S =0$, but  
we can see that $\Delta S = 0$ implies $\tau = \beta \hbar$. Second, the 
relation (\ref{deltas}) is adequate for an interpretation in the domain of 
restricted relativity. Since the entropy is considered as a relativistic 
invariant the right hand side of equation (\ref{deltas}) must be 
relativistic invariant. It appears as the scalar product of the the energy 
momentum tensor by a quantity $(\frac{1}{T}, (\frac{V}{T}))$ that must be a 
four-vector showing that $(\frac{1}{T})$ must behave as a time in a Lorentz 
transformation. This result also supports the idea that it exists a 
fundamental relation between time and the reverse of temperature.  \\  
The relation $\tau = \beta \hbar$ is extensively used in the derivation of 
the black hole properties but it appears more as a mathematical trick than a 
relation having a strong physical basis. In this domain it is usual to 
investigate the properties of analytic functions defined on a complex time 
plane. For a given spatial distance between two points it exists on the real 
time axis a given domain on which the interval is spacelike. On this domain 
the field commutator vanishes and this allows us to define a unique analytic 
continuation of the Green functions through the complex plane 
(\cite{fulling}). Working at a non-zero temperature the Green functions 
defined on the euclidean sector exhibits a periodicity $\tau = \beta \hbar$. 
To derive this result the crux is that the effect of temperature is 
determined via the thermal density matrix $\exp -\beta H$ in which $H$ is 
the hamiltonian operator. \\ In our derivation of the relation $\tau = \beta 
\hbar$ we do not need the hamiltonian operator nor the Schr\"{o}dinger 
equation or the canonical form of the density matrix. What we need from 
quantum mechanics is the existence of a relation that mimics the Heisenberg 
uncertainty relation and defines the spacetime fine structure. This is 
reminiscent of a strong result obtained by Wald (\cite{wald1}): some results 
of black hole theory can be derived without using the detailed expression of 
the Einstein equations. \\ From all arguments developped above it seems 
normal to conclude that the relation $\tau = \beta \hbar$ has a very general 
and deep physical meaning. 

\section{Relation between action and entropy}
In the Gibbs ensemble approach of statistical mechanics we cannot expect 
a relation between action and free energy because these two quantities have a 
very different nature: free energy is considered as an equilibrium quantity 
that we have to calculate by integration over the phase space while the 
action is a dynamical quantity and its definition requires the introduction 
of a time interval. In the previous Section we have developed a dynamical 
approach of thermodynamic equilibrium in which a time interval $\tau$ is 
associated with the reverse of the temperature. We have shown that the free 
energy defined according to $F = - k_{B}T \ln Z$ is a functional of an 
euclidean action $A^{E}[x(t);t,t_{0}]$ as in the case of black hole. Note 
that in our approach the zero loop approximation cannot be used without 
loosing the main part of the underlying physics therefore we will stay with 
the functional form. Before setting up a relation between action and entropy 
we analyze the relation between the lagrangian and euclidean version of the 
action. \\ In (\ref{upath}) we have defined a potential energy 
$<u_{P}>_{path}$ by an average over paths and the mean value of the kinetic 
energy over paths will be defined by $\frac{m}{2} <(\frac{\delta x }{\delta 
t})^{2}>_{path}$ that we have introduced in (\ref{kinpath}). As usually we 
may define a lagrangian by the difference between kinetic and potential 
energy. For an elementary time interval the corresponding action  $<A(T, 
\delta t)>_{path}$ will be \begin{equation} <A(T, \delta t)>_{path} = \delta 
t [\frac{m}{2} <(\frac{\delta x}{\delta t})^{2}>_{path}- <u_{P}>_{path}]= 
\frac {\hbar}{2} - [<u_{K}>_{path} +<u_{P}>_{path}]=  \frac {\hbar}{2} - U 
\delta t \label{action1} 
\end{equation} 
The second equality results from (\ref{kinpath}) and the third is the 
consequence of the equilibrium condition discussed in the previous Section. 
Thus from (\ref{action1}) we can see that the elementary lagrangian 
action is related to $- U \delta t$ and it seems natural of considering $U 
\delta t $ as the elementary euclidean action. In the limit $\delta t \to 
{0}$ the lagrangian action has a finite value $\frac{\hbar}{2}$ 
corresponding to the quantum of action, thus the action is not 
differentiable as expected. If we increase the temperature by $\delta T$ we 
have \begin{equation} <A(T + \delta T, \delta t)>_{path} -<A(T, \delta 
t)>_{path} = - \delta t [U(T+ \delta T) -U(T)] = - \delta t T \delta S 
\label{dA} \end{equation} in which the last equality is a consequence of the 
usual thermodynamic relation. The net change of action $\delta A_{path}$ on 
the time interval $\tau$ will be obtained by summing up $[<A(T + \delta T, 
\delta t)>_{path} -<A(T, \delta t)>_{path}]$ on all the elementary time 
interval covering the total time interval $\tau$. For the part of (\ref{dA}) 
containing $U(T)$ this will be simply done by multiplying the previous 
result by the number of elementary steps $i.e.$ $\frac{\tau}{\delta t}$. It 
is easy to see that the final result can be written as \begin{equation} 
\frac{\delta A_{path}}{\hbar} = - \frac{\delta S}{k_{B}} \label{dA2} 
\end{equation} For a black hole having an area $A$ the entropy is given by 
$S= \frac {k_{B}c^{3}}{4 G \hbar}A$ and the euclidean action is $A^{E} =- 
\frac{c^{3}}{4G} A$ leading to the relation $\frac{S}{k_{B}} = - 
\frac{A^{E}}{\hbar}$ from which we get immediatly (\ref{dA2}) provided we 
identify the change of $A_{path}$ with the change of $A^{E}$. This 
identification is justified since the introduction of an imaginary time in 
the lagrangian defined in (\ref{action1}) leads to minus the hamiltonian and 
therefore, in comparison with the usual definition an extra minus sign is 
introduced. It is important to note that in the case of black hole the first 
loop approximation leads to a direct relation between action and entropy but 
the relation (\ref{dA2}) is also verified. A relation like (\ref{dA2}) that 
introduces a link between variations fits quite well with the spirit of 
thermodynamics and we may think that it represents a result more general than 
$\frac{S}{k_{B}} = - \frac{A^{E}}{\hbar}$. 

\section{Concluding remarks}
Entropy is a fundamental quantity in physics and the law of evolution of 
entropy may be considered as the actual law of evolution for a system. 
But recent results show that, possibly, the concept of entropy is far to be 
well understood (\cite{wald1},\cite{marolf}). This is the case in the 
black hole domain in which we are not able to associate the entropy with a 
counting of microstates. In addition, as a consequence of the Unruh effect 
(\cite{marolf}), it appears that the number of microstates might be related 
to the motion of the observer. This suggests that a definition of entropy 
more general that the one introduced by Boltzmann might exist. This paper is 
an attempt to introduce a definition of entropy that we can use 
both for ordinary systems and systems including black hole. It is also an 
illustration showing that general results obtained in the domain of black 
hole can be extended to ordinary systems producing an improvement in our 
general understanding about thermodynamics. \\ We show that for ordinary 
systems the entropy can be calculated as in the case of black hole $i.e.$ by 
inspecting the properties of a spacetime model. Of course, starting from 
this new approach based on a dynamic description of the thermodynamic 
equilibrium we recover the standard results. But we also establish two 
results used in the domain of black hole physics. First, it exists a 
relation between time and temperature; this is much more than a mathematical 
trick as it may appear in the derivation of the KMS condition 
(\cite{fulling}). Here this relation is based on an equilibrium condition, 
it appears as very general since it is connected neither with the existence 
of a canonical density matrix nor with the existence of the Schr\"{o}dinger 
equation. Second, we show that it exists a relation between action and 
entropy that is also verified in the case of black hole. \\  
It is interesting to note that in our approach we do not follow 
the traditional route going from the Schr\"{o}dinger equation to the 
calculation of eigen states for energy and then to the estimation of the 
partition function via the density matrix. In fact we follow the opposite 
route. We start from a spacetime model in which the quantification appears 
via a relation that determines the spacetime fine structure and consider 
directly the paths in spacetime, the motion is then characterize by a 
positive semi-group but not by a unitary transformation. However, if we 
force the system to have a time-reversible behavior then, as shown in 
(\cite{jpb1}), we may recover the Schr\"{o}dinger equation and 
an unitary law of evolution. \\
Finally, the results obtained in this paper
suggest the following conjecture: it is possible that the general definition 
of entropy is connected with a definition of the order in spacetime 
rather than associated with a counting of microstates.


\end{document}